\RequirePackage{graphicx}

\documentclass{iau}

\usepackage{amsmath}
\usepackage{graphicx}
\usepackage{multirow}

\usepackage{siunitx}
\usepackage[capitalize]{cleveref}

\usepackage{./aasmacros}
\DeclareSIUnit\parsec{pc}
\DeclareSIUnit\year{yr}
\DeclareSIUnit\years{yrs}
\DeclareSIUnit\dex{dex}
\DeclareSIUnit\mag{mag}
\DeclareSIUnit\h{\mathnormal{h}}
\DeclareSIUnit\Msun{M_\odot}
\DeclareSIUnit\Rsun{R_\odot}
\DeclareSIUnit\Lsun{L_\odot}

\newcommand*{\OIII}{[O\,\textsc{iii}]} % [OIII]

\begin{document}

\lefttitle{Lucas M.\ Valenzuela et al.}
\righttitle{Deep imaging meets motion}

\journaltitle{The Hidden Beauty of the Galactic Outskirts}
%\jnlPage{1}{7}
\jnlDoiYr{2025}
\doival{10.1017/xxxxx}
\volno{403}

\aopheadtitle{Proceedings IAU Symposium}
\editors{D. Martínez-Delgado}
 
\title{Deep Imaging Meets Motion: \\ Complementing Stream Photometry Through Planetary Nebula Kinematics}

\author{Lucas M. Valenzuela$^1$, Johannes Stoiber$^1$, and Rhea-Silvia Remus$^{1,2}$}
\affiliation{$^1$Universitäts-Sternwarte, Fakultät für Physik, Ludwig-Maximilians-Universität München, Scheinerstr. 1, 81679 München, Germany}
\affiliation{$^2$Centre for Astrophysics and Supercomputing, Swinburne University of Technology, Hawthorn VIC 3122, Australia}

\begin{abstract}
The combination of deep imaging data and kinematic measurements in galaxy outskirts promises to reveal extensive insights into the structure and history of individual galaxies.
From a census of tidal features around galaxies from the Magneticum simulation, we disentangle the dynamics for a selected stellar stream from the underlying halo by identifying the stream progenitor galaxy.
While these dynamics are challenging to measure observationally, we show that they are effectively obtained through planetary nebulae (PNe) as tracers, which we model in the simulation using the PN framework PICS (PNe In Cosmological Simulations).
We find that the PNe in the brightest \SI{1.5}{\mag} of their luminosity function are sufficient to recover the underlying stellar dynamics of the massive stream.
We thereby establish PNe as an attractive alternative to expensive deep IFU observations, where combining low-surface-brightness observations and PN dynamical measurements will enhance our ability to constrain the gravitational potential of galaxies.
\end{abstract}


\maketitle

\section{Introduction}

Current and upcoming wide-field deep imaging facilities like Euclid, Rubin, and Roman are delivering vast amounts of low-surface-brightness data in the outskirts of galaxies \citep[e.g.,][]{trujillo+21, euclid_collaboration+22:XVI, martin+22}.
These outer structures hold clues to both the formation history and the underlying gravitational potential of the host galaxy \citep[e.g.,][]{hendel&johnston15, mancillas+19, pearson+22}.
Extragalactic tidal streams and shells are the extended remnants of tidally disrupted galaxies that are orbiting a more massive host galaxy, where streams tend to arise from minor mergers on more circular orbits, whereas shells are the result of radial orbits \citep[e.g.,][]{karademir+19}.
For this reason, these structures are useful in constraining the host galaxy's potential.
In particular, stellar streams have been used to this end by tracing their morphology and reconstructing their orbit, not only in the Milky Way from 6D phase space information \citep[e.g.,][]{koposov+23, starkman+23}, but also extragalactically \citep[e.g.,][]{walder+25}.

On the extragalactic front, the projected view of faint integrated stellar light complicates the process of modeling the gravitational potential due to the limited information available.
Pure photometric information provides the means to constrain the form of the potential, but a further velocity measurement is necessary to obtain absolute values of scale radius and enclosed total mass \citep[e.g.,][]{pearson+22, walder+25}.
Line-of-sight velocity measurements of faint structures are currently possible for individual tidal features \citep[e.g.,][]{fensch+20}, but this will not be feasible for the wealth of upcoming deep photometric observations of galaxy outskirts.
As a possible solution, kinematic tracer populations such as planetary nebulae (PNe) could be employed as they have long been used to trace the low-surface-brightness Universe given their bright emission in \OIII{} $\lambda5007$ \citep[e.g.,][]{arnaboldi+94, mendez+09}.

Hydrodynamical cosmological simulations have proven to be useful tools to study and interpret the occurrence of low-surface-brightness structures \citep[e.g.,][]{pop+18, mancillas+19, valenzuela&remus24, khalid+25, stoiber+25}.
Additionally, PN models have made significant advances in recent years, developing new approaches to reproduce the universal PN luminosity function (PNLF) with its bright-end cutoff at \SI{-4.5}{\mag}, for which the theoretical understanding had long been limited \citep[e.g.,][]{ciardullo10, gesicki+18, valenzuela+25:picsI}.

In this work, we present an application of the new PN model PICS \citep{valenzuela+25:picsI} to the simulation Magneticum Pathfinder \citep{dolag+25:magneticum} to predict the effectiveness of PNe in tracing the dynamics of a stream for a first case study.
In \cref{sec:simulation_model}, we introduce the simulation and tidal feature sample from which the stream is selected as well as the PICS framework.
In \cref{sec:results}, we analyze the galaxy halo and stream kinematics, for which we then demonstrate how well the PNe trace them.
Finally, we summarize and conclude our findings in \cref{sec:conclusion}.

\section{Simulation and Model}
\label{sec:simulation_model}

To investigate the usefulness of kinematic tracer populations for tidal features, cosmological simulations with a sufficient resolution are needed as they contain galaxies with self-consistent star formation histories in a cosmological context.
In this section, we first introduce the cosmological simulation used in this work and the census of tidal features found therein, followed by a description of the PN model that we applied to the stellar component.

\subsection{Magneticum Pathfinder Simulation}
\label{sec:magneticum}

The Magneticum Pathfinder simulation suite \citep{dolag+25:magneticum} includes simulations ranging from large cosmologically relevant volumes down to smaller high-resolution box sizes useful for galaxy studies.
For the investigation of tidal features, we use Magneticum Box4 (uhr), which has a side length of \SI{68}{\mega\parsec} and was initialized with $2 \times 576^3$ DM and gas particles (stellar particle masses $m_* = \SI{1.3e6}{\Msun\per\h}$ and softening length $\epsilon_* = \SI{1}{\kilo\parsec}$).
It was run with an extended version of \textsc{Gadget}-2 \citep{springel05:gadget2} with smoothed-particle hydrodynamics (SPH). For details about the implemented physics, refer to \citet{dolag+25:magneticum}.
The galaxies were identified with \textsc{SubFind}, adapted for baryonic simulations \citep{springel+01:subfind, dolag+09:subfind}.

The galaxies and associated scaling relations in Box4 have been shown to agree well with observations \citep{dolag+25:magneticum}, including their outer kinematics \citep[e.g.,][]{schulze+20:II}, dynamics \citep[e.g.,][]{harris+20}, and matter distribution and in situ component fractions \citep[e.g.,][]{remus&forbes22}.
For the simulated galaxies, we have created a census of tidal features around the galaxies in previous work and showed that their occurrence as well as connection to the inner galaxy kinematics is consistent with observations from MATLAS \citep{valenzuela&remus24}, with subsequent work finding good agreement with the SAMI survey \citep{rutherford+24}.
In a follow-up study, we compared the sizes and positions of streams and shells between the simulation and several observations, again finding good agreement \citep{stoiber+25}.
As a result, the Magneticum Box4 (uhr) simulation is an ideal simulation for the purposes of studying tidal features and their dynamics in closer detail.

\subsection{Tidal Feature Census}
\label{sec:tidal_features}

\begin{figure}
    \centering
    \includegraphics[width=0.48\linewidth]{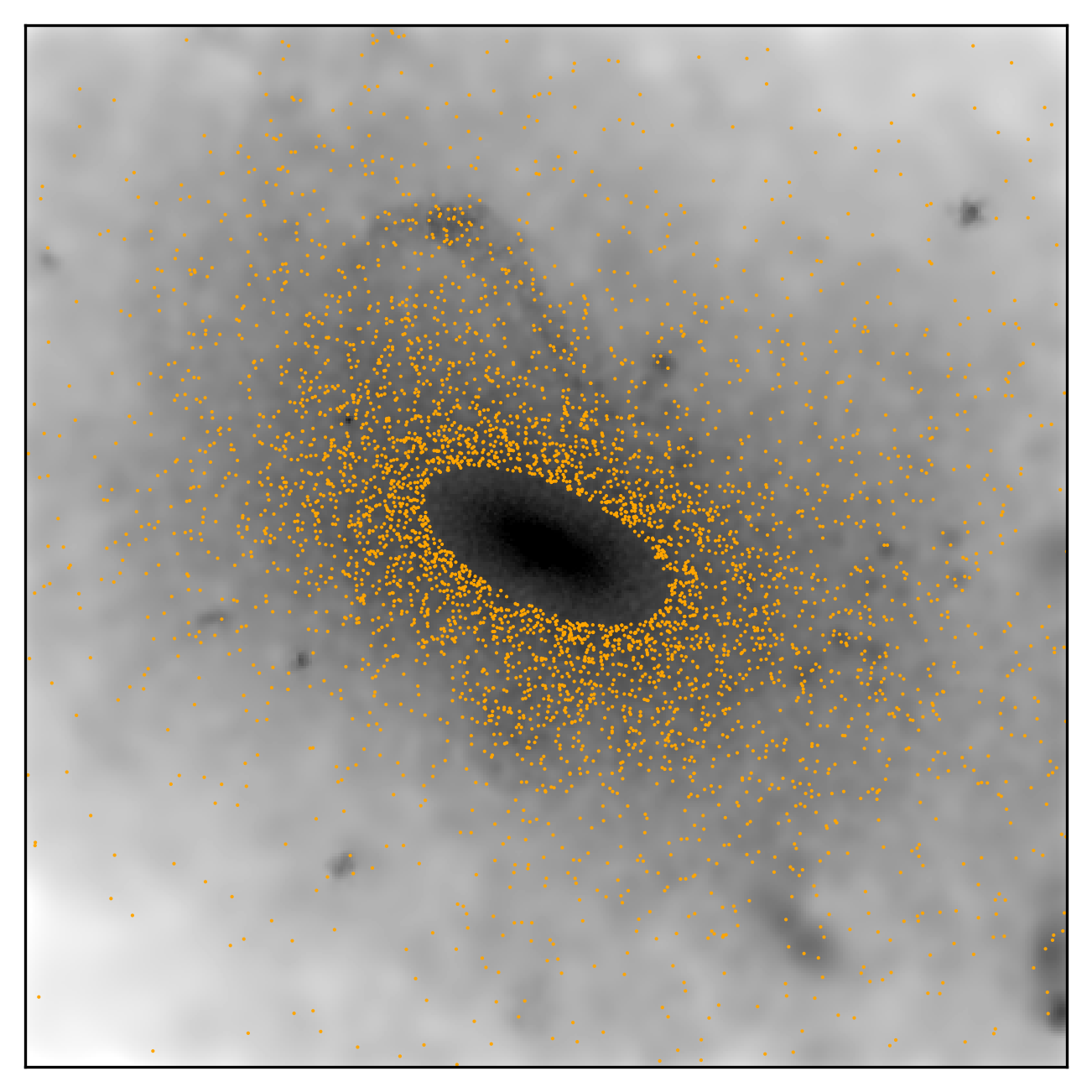}
    \includegraphics[width=0.48\linewidth]{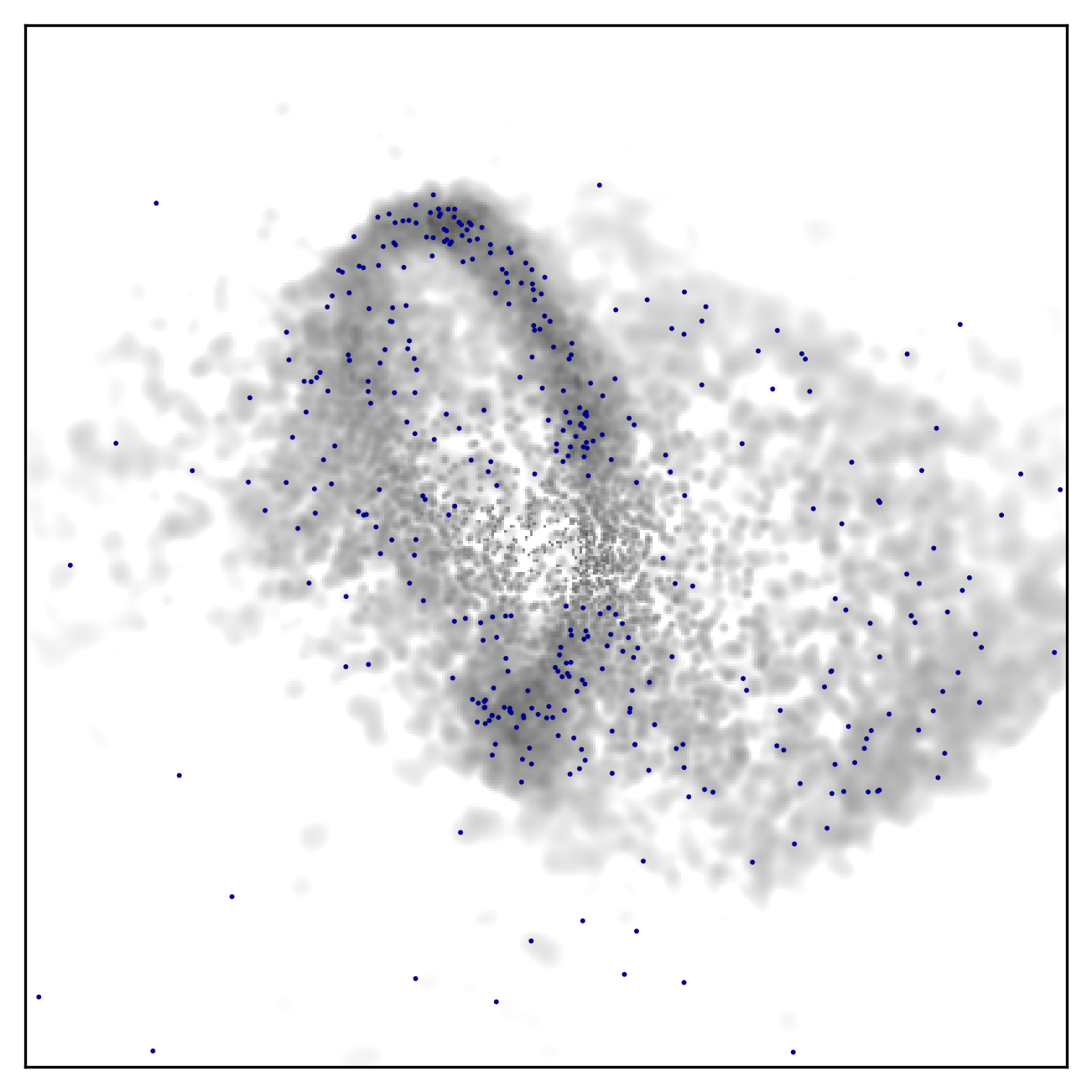}
    \caption{
        \textit{Right}: Mock observation of a prominent stream to the upper left of its host galaxy from Magneticum Box4 in the r-band with overplotted PNe modeled with PICS (orange points) in the outer galaxy regions. The galaxy is viewed from a random projection.
        \textit{Left:} Stream separated from the host galaxy, as identified by the stellar particles originating from the stream progenitor galaxy. The PNe associated with these stellar particles are overplotted in blue.
        The PNe shown in both panels are those brighter than the threshold absolute magnitude of $M(5007) \leq \SI{-2.0}{\mag}$, which corresponds to the PNe in the brightest $\SI{2.5}{\mag}$ below the bright-end cutoff.
        Both panels have a side-length of \SI{140}{\kilo\parsec} and cover the same area.
    }
    \label{fig:streamgal}
\end{figure}

For the tidal feature census by \citet{valenzuela&remus24}, streams, shells, and tails were visually identified around 520~galaxies with stellar masses of $M_* \geq \SI{2.4e10}{\Msun}$ and half-mass radii of $R_{1/2} > \SI{2}{\kilo\parsec}$.
The classification resulted in 62~galaxies with streams, 21~with shells, and 26~with tails, where the fraction of galaxies with a given type of feature increases with stellar mass, just as in observations \citep[see fig.~3 of][]{valenzuela&remus24}.
By tracing the stellar particles of each tidal feature back in time, we additionally identified the respective progenitor galaxies, enabling us to separate the present-day features from the galaxy and halo \citep{stoiber+25}.
This is particularly important in the context of this work for separating the feature dynamics from that of the halo.
An example galaxy with a prominent stream is shown in the left panel of \cref{fig:streamgal}, which we use in this work for the case study.
The host galaxy has a mass of $M_* = \SI{4e11}{\Msun}$ and a projected half-mass radius of $R_{1/2} = \SI{13.5}{\kilo\parsec}$, while the stream originates from a galaxy with $M_* = \SI{2.4e10}{\Msun}$.
The stream wraps around the galaxy, with a still identifiable core located at the top of the stream.

\subsection{The PICS Model}
\label{sec:pics}

We obtain the PN populations from the PICS  framework \citep[PNe In Cosmological Simulations;][]{valenzuela+25:iau, valenzuela+25:picsI}, which models the relevant physical processes for PNe given a stellar particle that represents a single stellar population.
Stellar evolution including the post-asymptotic giant branch (post-AGB) phase and initial-to-final mass relation is followed through the metallicity-dependent models by \citet{miller_bertolami16}.
The nebular \OIII{} emission is determined through an empirical nebular model \citep{valenzuela+19} with a metallicity correction motivated from theory \citep{dopita+92}.
Finally, circumnebular extinction is accounted for through an empirical relation from recent observations \citep{jacoby&ciardullo25, valenzuela+25:picsII}.
The result is a population of PNe with their respective \OIII{} magnitudes.
The model has already been successfully used to interpret observations of elliptical galaxies and their PN populations \citep{soemitro+25}.

Applying PICS to the selected galaxy with the prominent stream reveals a PN population that largely follows the underlying stellar component in spatial distribution.
Because the PNe are modeled for the individual stellar particles, we were also able to separate those belonging to the stream from those of the host galaxy.
The orange and blue points in \cref{fig:streamgal} show the PNe in the brightest $\SI{2.5}{\mag}$ of the PNLF, where for old elliptical galaxies a luminosity-specific PN number on the order of around $\alpha_{2.5} \sim \SI{e-8}{\per\Lsun}$ is expected from observations (i.e., the number of PNe per underlying bolometric luminosity; e.g., \citealp{hartke+22:II}).

\section{Tracing Tidal Stream Dynamics}
\label{sec:results}

Given the identified tidal features with their positions and spatial distributions, it is possible to model the potential of the host galaxy.
In particular, there are significant efforts for this based on streams in the outskirts \citep[e.g.,][]{pearson+22, walder+25}.
However, because of uncertainties in distance to the respective systems, the constraints are generally limited to a relative description of the density profile.
To determine the absolute mass of the host halo, a single line-of-sight velocity measurement of a given stream is in fact sufficient.
In the following, we present the stellar kinematics of the example galaxy as well as of its tidal stream and to what extent it is traced by the PNe.

\subsection{Dynamical decomposition of the stellar component}

\begin{figure}
    \centering
    \includegraphics[width=0.48\linewidth]{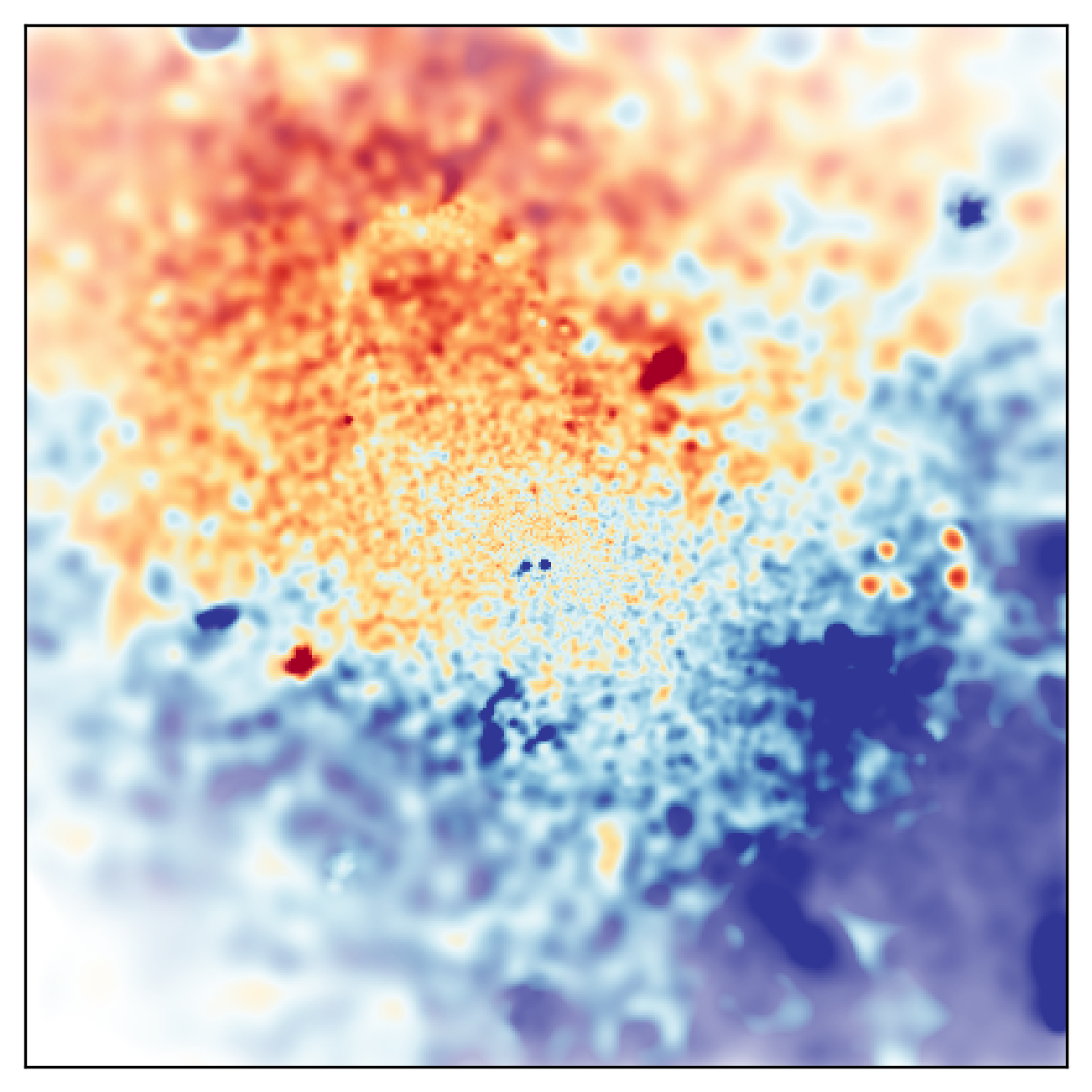}
    \includegraphics[width=0.48\linewidth]{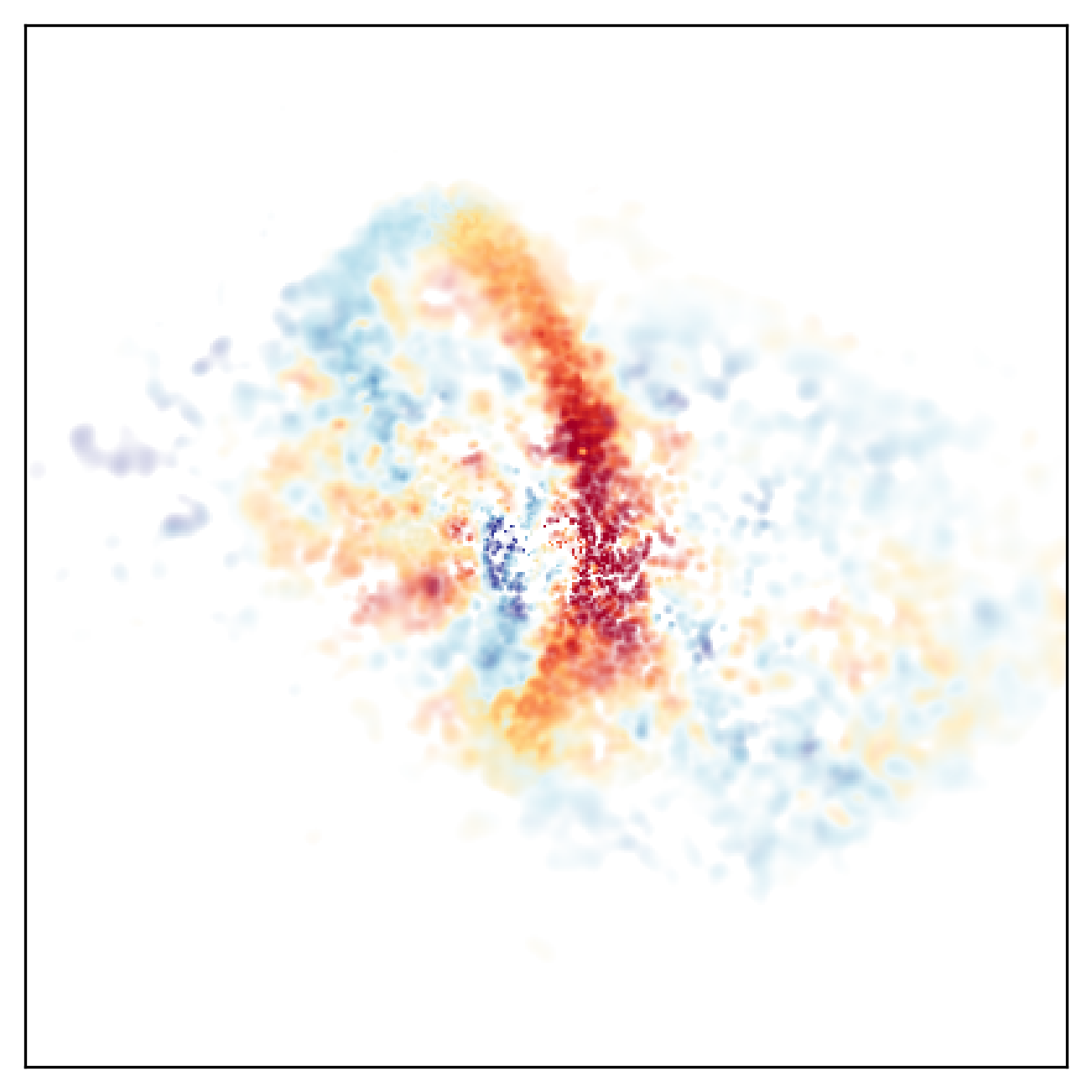}
    \caption{
        Line-of-sight stellar velocity maps of the same galaxy and stream as in \cref{fig:streamgal} with the same projection and area displayed. The color map ranges from \num{-150} to \SI{150}{\kilo\meter\per\second} for the dark red and dark blue extremes, respectively. The color opacity is furthermore set by the r-band flux shown in \cref{fig:streamgal}, with fainter regions appearing more white in the velocity maps.
        \textit{Right}: Velocity map of the entire galaxy including the stream.
        \textit{Left}: Velocity map of the stream stellar particles.
        For the visualizations, each of the stellar particles is smoothed according to the local density, after which the displayed line-of-sight velocities in each pixel are determined by the light-weighted mean.
    }
    \label{fig:streamgal_velmap}
\end{figure}

To this end, we show the line-of-sight stellar velocities as they would be observed for the entire galaxy in \cref{fig:streamgal_velmap} (left panel), as well as the line-of-sight velocities of the separated stream stellar particles (right panel).
The halo shows a clear sign of ordered rotation around a rotational axis drawn from the top right to the bottom left.
Interestingly, this is slightly misaligned with respect to the inner galaxy shape and its position angle (PA), which is generally related to the merger history \citep[e.g.,][]{foster+16, schulze+20:II, valenzuela+24:shapesI}.
The stream kinematics are hardly visible in the full galaxy velocity map, where only a slight drop in velocity can be seen at the top right of the galaxy (left panel), which corresponds to the location where the stream is best visible in the photometric map (left panel of \cref{fig:streamgal}).
The pure stream kinematics primarily show how the stream follows an orbit around the host, but also to what extent phase space mixing has already made the strongly stripped stellar component (seen to the very right in the maps) nearly indistinguishable from the rest of the halo component (comparing the right-most region between the left and right panels).

\subsection{Planetary nebulae as kinematic tracers for tidal features}

\begin{figure}
    \centering
    \includegraphics[width=0.37\linewidth]{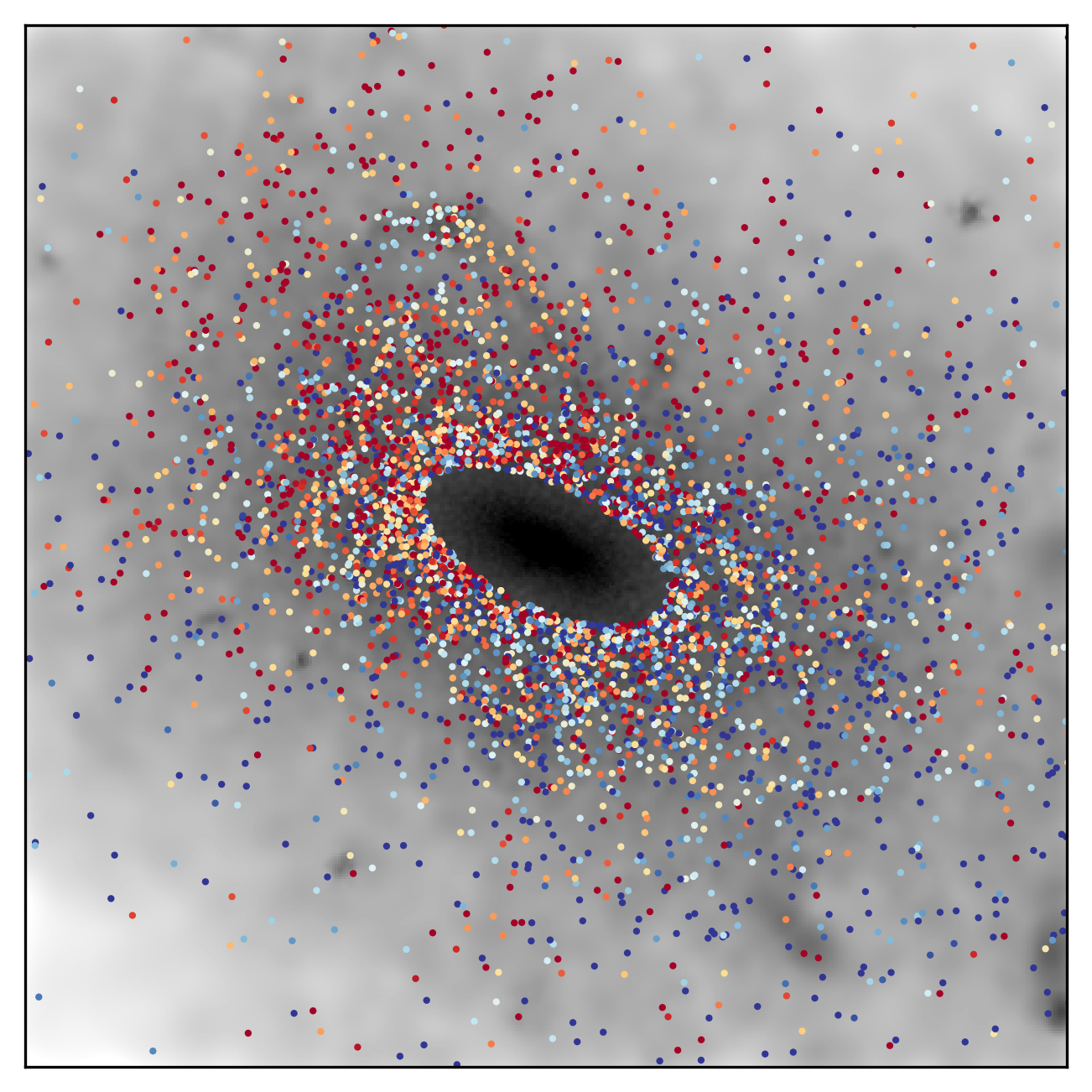}
    \includegraphics[width=0.37\linewidth]{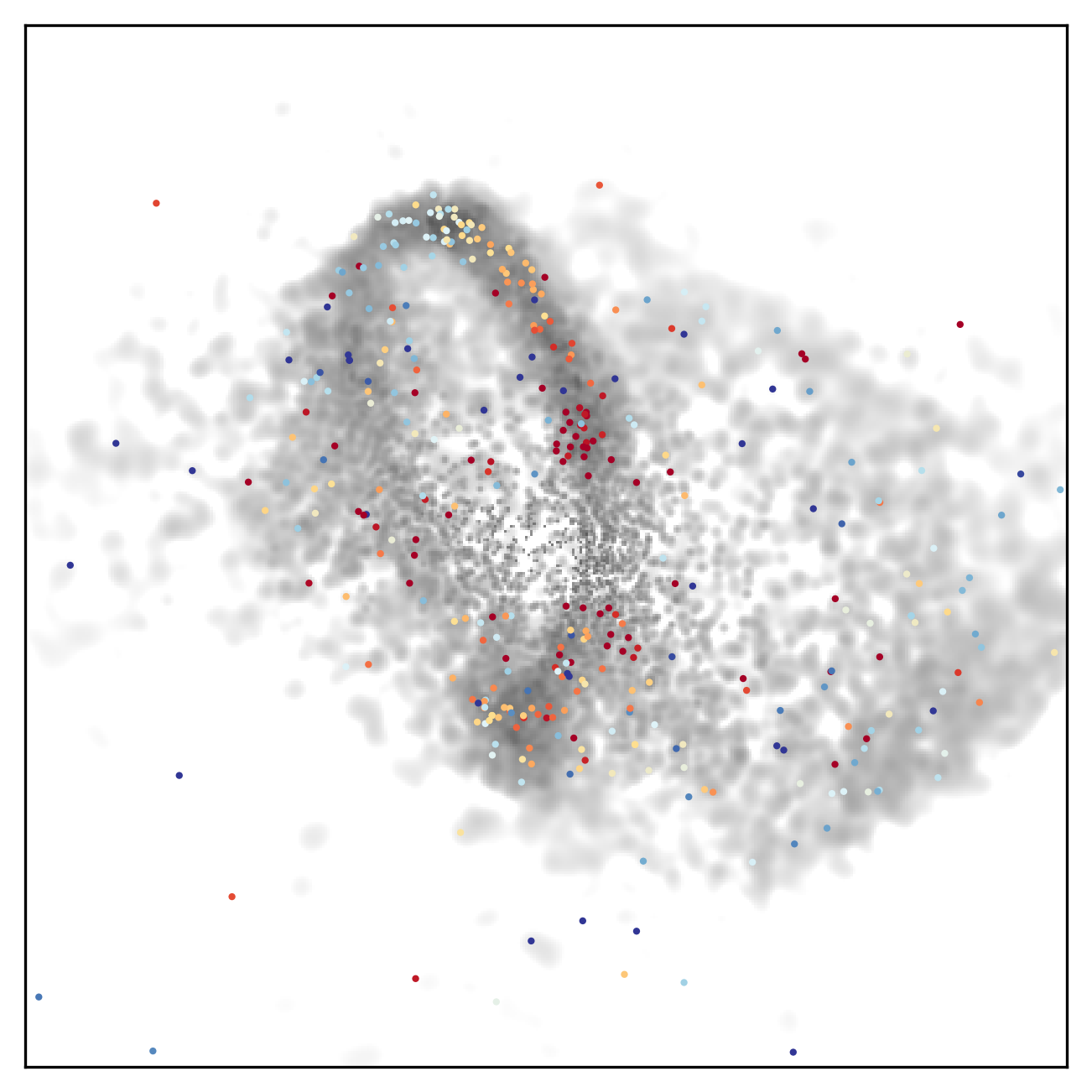}
    \includegraphics[width=0.36\linewidth]{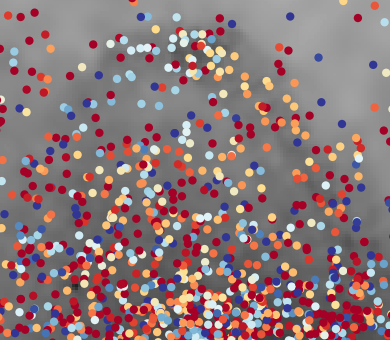}
    \includegraphics[width=0.36\linewidth]{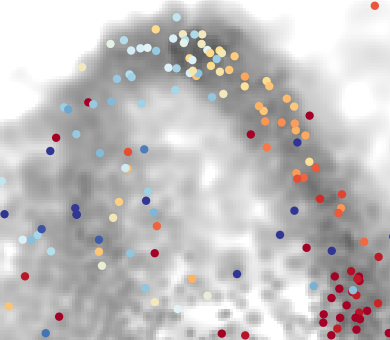}
    \includegraphics[width=0.36\linewidth]{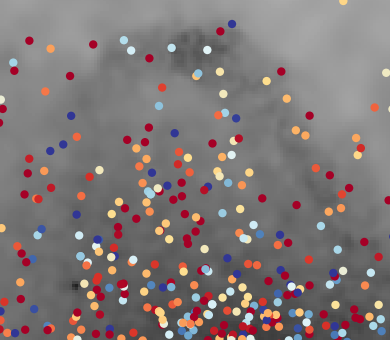}
    \includegraphics[width=0.36\linewidth]{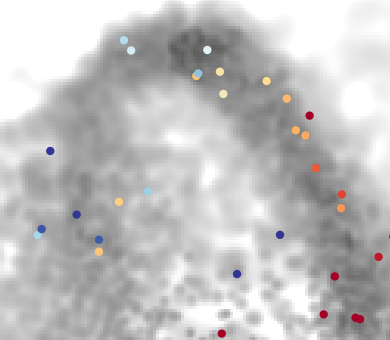}
    \includegraphics[width=0.36\linewidth]{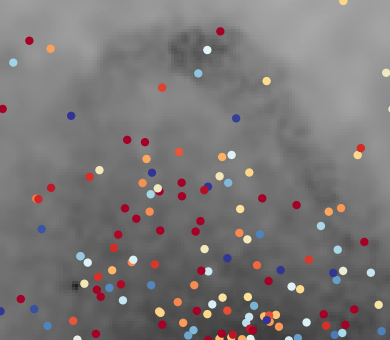}
    \includegraphics[width=0.36\linewidth]{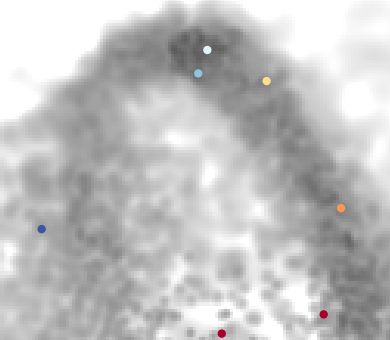}
    \caption{
        Mock observations of the same galaxy and stream as in \cref{fig:streamgal,fig:streamgal_velmap} with overplotted PNe colored by their line-of-sight velocities. The color map ranges from \num{-150} to \SI{150}{\kilo\meter\per\second} for the dark red and dark blue extremes, respectively.
        \textit{Top row}: Same projection and area as shown in \cref{fig:streamgal,fig:streamgal_velmap}, where only PNe are displayed that are brighter than a magnitude limit of $M(5007) \leq \SI{-2.0}{\mag}$, just as in \cref{fig:streamgal}.
        \textit{Bottom rows}: Zoomed-in region of the bright upper stream part, where the overplotted PNe are subject to magnitude cuts of $M(5007) \leq -2.0$, \num{-3.0}, and \SI{-3.5}{\mag}, from top to bottom. These correspond to the brightest 2.5, 1.5, and \SI{1.0}{\mag} of the PNLF, respectively.
        The zoomed-in region has a width of around \SI{36.5}{\kilo\parsec}.
    }
    \label{fig:streamgal_velmap_pne}
\end{figure}

While only one kinematic measurement along the stream would suffice to anchor the halo mass in absolute terms, the question arises to what extent it is possible to obtain this from individual PNe without knowing whether they belong to the stream or halo.
In \cref{fig:streamgal_velmap_pne}, we show the same PN population as in \cref{fig:streamgal}, but colored by line-of-sight velocity (top row).
The PN kinematics overall show the same behavior as the stellar kinematics in \cref{fig:streamgal_velmap}, that is, they clearly trace the halo's global rotation.
Given that PNe trace the stellar component, this was to be expected.
Additionally, now the individual line-of-sight velocities of the PNe simultaneously provide some insight into the underlying velocity dispersion, where in particular in the inner regions there are strongly varying PN line-of-sight velocities found close together.

The following rows show the zoomed-in region of the bright upper stream part with the overplotted PN population colored by velocity.
They show the PNe with varying magnitude limits per row where the first of those rows shows the largest magnitude range of $M(5007) \leq \SI{-2.0}{\mag}$ (same as the top row of \cref{fig:streamgal_velmap_pne}), followed by \num{-3.0} and \SI{-3.5}{\mag}.
The decreasing number of PNe with each row raises the stochasticity of whether it is possible to detect a sufficient number of PNe associated with the stream to obtain the necessary corresponding velocity measurement.
While the stream kinematics are clearly traced by the brightest \SI{2.5}{\mag} PNe (right panel in the second row of \cref{fig:streamgal_velmap_pne}, $M(5007) \leq \SI{-2.0}{\mag}$), there are few PNe in the brightest \SI{1}{\mag} (fourth row) with only five PNe lying along the main stream section seen on the right side of the right zoom-in panel.
The additional halo PNe seen within the stream in the left zoom-in panel of the fourth row hinder a precise estimate of the stream dynamics.

For the intermediate deep PN sample of the brightest \SI{1.5}{\mag} (third row of \cref{fig:streamgal_velmap_pne}, $M(5007) \leq \SI{-3.0}{\mag}$), the stream PNe are sufficient to trace the stream dynamics well through their line-of-sight velocities (right panel).
In particular, the smooth velocity gradient along the brightest stream section is recovered, from light blue points at the top to dark red points at the bottom right.
Further including the halo PNe (left panel), the PNe found within the projected brightest stream section still largely trace the stream dynamics, with only some contaminating outliers.
The total stream mass displayed in the zoomed-in panels of the right column is $M_{*,\text{zoom-stream}} = \SI{6.7e9}{\Msun}$, which corresponds to between a third and a fourth of the original progenitor galaxy's mass.

This means that for a massive stream such as in the case study here, detecting the brightest \SI{1.5}{\mag} PNe is sufficient to obtain a robust dynamic measurement of the stream along the brightest section of the stream.
This method even promises to deliver a velocity gradient along the stream, which is expected to further improve the constraints that can be placed on the absolute halo mass of the host galaxy with methods such as those employed by \citet{pearson+22} and \citet{walder+25}.

\section{Summary and Conclusion}
\label{sec:conclusion}

In this work, we have applied the PICS framework for modeling PNe \citep{valenzuela+25:picsI} to the cosmological simulation Magneticum Box4 (uhr; \citealp{dolag+25:magneticum}) to investigate to what extent PN kinematics can trace the dynamics of tidal streams.
We present a case study for a prominent tidal stream around a massive elliptical galaxy, where the line-of-sight stellar kinematics along the stream show a clear velocity gradient.
By applying different brightness cutoffs for the PNe with respect to their bright \OIII{} emission line, we analyze to what extent the stream can be disentangled from the underlying halo kinematics solely based on the PN line-of-sight velocities and positions.
Over the brightest stream region with a mass of \SI{6.7e9}{\Msun}, the brightest \SI{1.5}{\mag} PNe are sufficient to recover the stream dynamics within the host galaxy's halo.

As a result, we show that PNe are effective tracers of the underlying stellar kinematics, even for tidal streams when they are massive enough.
Moving towards lower-mass tidal features, deeper observations will be required to obtain a sufficient sample of PNe.
We find that tracer populations are an attractive alternative to expensive low-surface-brightness kinematic measurements for obtaining the velocity data, which are needed to recover the halo profile and mass from stream detections \citep[e.g.,][]{pearson+22, walder+25}.
In the era of wide-field deep photometric surveys, where tidal features are being discovered around increasingly large samples of galaxies \citep[e.g.,][]{martinez_delgado+23, gordon+24}, PN surveys targeting the tidal features are a viable approach to complement the photometry.
In particular, wide-field narrow-band imaging can be used to identify PN candidates.
Later spectroscopic follow-up measurements could then deliver the relevant kinematic information.
We conclude that PN kinematics are a feasible way forward for delivering the much-needed complementary information to the low-surface-brightness imaging of galaxy outskirts from the observational side.

\section*{Acknowledgments}
LMV acknowledges support by the German Academic Scholarship Foundation (Studienstiftung des deutschen Volkes) and the Marianne-Plehn-Program of the Elite Network of Bavaria.
JS acknowledges support by the COMPLEX project from the European Research Council (ERC) under the European Union’s Horizon 2020 research and innovation program grant agreement ERC-2019-AdG 882679.
The \textsc{Magneticum Pathfinder} simulations were performed at the Leibniz-Rechenzentrum with CPU time assigned to the Project pr83li.

\bibliographystyle{mnras}
\bibliography{bib}

\end{document}